\documentclass[10pt, twocolumn, letterpaper]{article}

\usepackage[pagenumbers]{iccv} %

\newcommand{\TODO}[1]{\textbf{\color{red}[TODO: #1]}}
\newcommand{\MEMO}[1]{\textbf{\color{blue}[MEMO: #1]}}

\renewcommand{\TODO}[1]{}

\renewcommand{\MEMO}[1]{}

\definecolor{iccvblue}{rgb}{0.21,0.49,0.74}
\usepackage[pagebackref, breaklinks, colorlinks, allcolors=iccvblue]{hyperref}

\usepackage{bm}
\usepackage{amsmath,amssymb}
\usepackage{url}
\usepackage{multirow}
\usepackage{tikz}
\usepackage{comment}
\usepackage[accsupp]{axessibility}  %
\usepackage{etoolbox}

\usepackage[capitalize]{cleveref}
\crefname{appendix}{Sec.}{Secs.}
\Crefname{appendix}{Sec.}{Secs.}

\newlength\squareheight
\usepackage{algorithm}
\usepackage{listings}
\usepackage{algorithmic}

\newcommand{\suppref}[1]{Supp.~\cref{#1}}
\newcommand{\ours}[0]{LayerD}

\def\paperID{9379} %
\def\confName{ICCV}
\def\confYear{2025}

\title{\ours{}: Decomposing Raster Graphic Designs into Layers}

\author{
    Tomoyuki Suzuki\textsuperscript{1} \quad Kang-Jun Liu\textsuperscript{2} \quad Naoto Inoue\textsuperscript{1} \quad Kota Yamaguchi\textsuperscript{1} \\
    $^{1}$CyberAgent \quad $^{2}$Tohoku University 
}

\begin{document}

    \twocolumn[
    \begin{@twocolumnfalse}
    {%
    \maketitle
    \vspace{-3mm}
    \centering
    \includegraphics[keepaspectratio, width=1\linewidth]{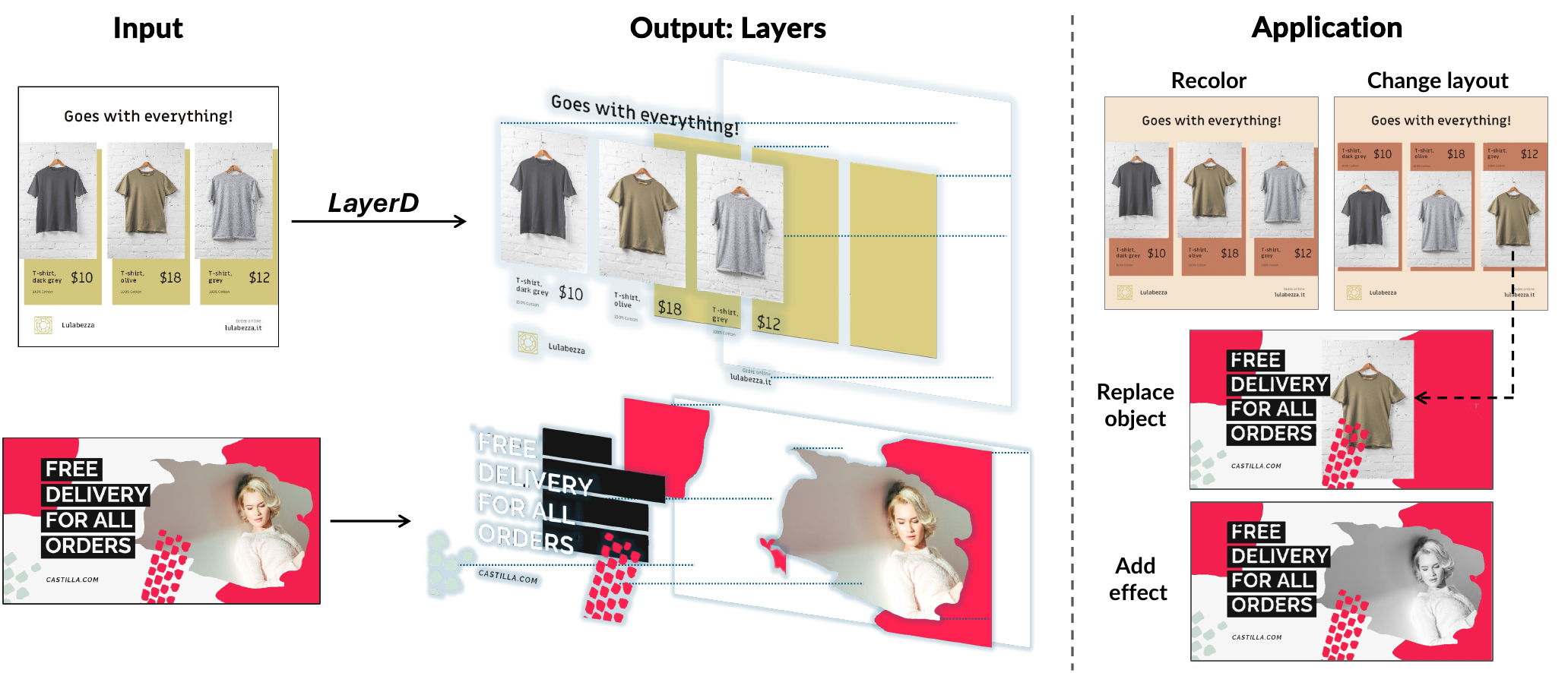}
    \captionof{figure}{
      \ours{} effectively decomposes raster graphic design images into layers, where the input design contains various elements such as typographic entities, embellishments, vector shapes, or even image materials.
      Once decomposed, one can apply image editing operations such as color conversion or translation at the layer level, or further apply other post-processing such as OCR to vectorize each raster layer.
    }
    \label{fig:teaser}
    \vspace{3mm}
    }
    \end{@twocolumnfalse}]
    \begin{abstract}
Designers craft and edit graphic designs in a layer representation, but layer-based editing becomes impossible once composited into a raster image. In this work, we propose \ours{}, a method to decompose raster graphic designs into layers for re-editable creative workflow. \ours{} addresses the decomposition task by iteratively extracting unoccluded foreground layers. We propose a simple yet effective refinement approach taking advantage of the assumption that layers often exhibit uniform appearance in graphic designs. As decomposition is ill-posed and the ground-truth layer structure may not be reliable, we develop a quality metric that addresses the difficulty. In experiments, we show that \ours{} successfully achieves high-quality decomposition and outperforms baselines. We also demonstrate the use of \ours{} with state-of-the-art image generators and layer-based editing. Code and models are publicly available \footnote{\url{https://cyberagentailab.github.io/LayerD/}}.
\end{abstract}

    \section{Introduction}
\label{sec:intro}
In the creative workflow, designers create and edit graphic designs at the \emph{layer} level, which is a basic unit of visual objects, such as text or images, and is commonly seen in design authoring tools like Adobe Photoshop or PowerPoint.
Once the workflow is complete, authoring tools composite these layers into a final image and deliver it to a display device or print media, such as social media posts, flyers, and posters.
Composite raster images do not retain layer information, making it difficult for designers to edit or retouch a raster graphic design.
Precise decomposition of raster artwork into layers, \ie, the inverse problem of composition, addresses this situation and enables a workflow that uses existing raster artwork assets to create new artwork.

In this work, we investigate graphic layer decomposition, aiming to automatically decompose a raster graphic design into a composable sequence of raster layers.
Since designers create graphic designs in a layered format, we can view this task as restoring the original layered representation.
Layer decomposition involves several computer vision tasks, such as object localization, segmentation, order estimation, and image inpainting.
Unlike natural images, graphic design is a mixture of various elements, including typography, embellishments, vector art, illustrations, and even natural image materials (\cref{fig:teaser}).
Naively applying image decomposition approaches~\cite{zhan2020self,zheng2021visiting,mulan} tuned for the natural image domain results in unintended decomposition (\eg, objects in a photo material are decomposed) or undesirable artifacts (\eg, background lighting affects solid-color vector-art), which are prohibitive for creative work.
Graphic layer decomposition is also inherently ill-posed; there are multiple possible solutions, and a layer can be arbitrarily divided into multiple layers.
This can be problematic, particularly when ensuring consistent evaluation.

We propose a method for fully automatic graphic layer decomposition, {\bf \ours{}}, which we formulate as iterative \emph{top-layer} matting and background completion.
We define a top-layer by objects appearing on the front without occlusion in the raster image, and in graphic designs, they typically contain typography at the beginning, followed by embellishment behind texts or photo materials in later iterations.
We learn a top-layer matting model from a high-quality graphic design dataset to ensure that the layer granularity aligns with humans, and together with an off-the-shelf inpainting model and a simple-yet-effective heuristic refinement to remove artifacts, we build a complete layer decomposition pipeline for graphic designs.
There have been a few similar attempts at fully automatic image decomposition into layer representations~\cite{mulan,accordion}, where they build a modular decomposition pipeline consisting of components for each subproblem, such as object detection~\cite{yao2023detclipv2,liu2023visual}, segmentation~\cite{ravi2024sam}, ordering~\cite{ranftl2020towards,lee2022instance}, and inpainting~\cite{rombach2022high}.
While the stacked pipeline approach can take advantage of pre-trained models at each stage, component stacking cannot avoid error accumulation throughout the pipeline; \eg, segmentation can fail when object detection contains overlapping bounding boxes or there is a large hall in the region.
\ours{} unifies detection, segmentation, and layer ordering by an iterative matting model to reduce the error accumulation while improving the efficiency.
In addition, we introduce a refinement approach for both foreground and background layers utilizing the domain prior that graphic design often consists of texture-less flat regions, which improves the final decomposition quality.

As layer decomposition can have multiple solutions and even humans are not consistent on the granularity of layers, we propose qualitative metrics for evaluation based on edit distance and visual quality between layer sequences aligned by dynamic time warping (DTW)~\cite{Mueller07_InformationRetrieval_SPRINGER}, which account for the inconsistency of the ground truth layers.
We compare \ours{} with several baselines and demonstrate that our method achieves the highest quality.

We summarize our contributions as follows:
\begin{itemize}
    \item We propose \ours{}, a fully automatic framework for layer decomposition from raster graphic designs. \ours{} unifies the subtasks inherent in layer decomposition into iterative top-layer extraction and leverages domain priors to improve the final decomposition quality.
    \item We propose a consistent evaluation protocol for layer decomposition based on the edit distance and appearance quality between aligned layer sequences, which accounts for the ambiguity in the ground truth layer structure.
    \item We empirically show that \ours{} achieves the highest quality compared to baselines and decomposed layers can be used for downstream graphic design editing.
\end{itemize}

    \section{Related Work}
\label{sec:related_works}

\begin{figure*}[t]
    \centering
    \includegraphics[keepaspectratio, width=1\linewidth]{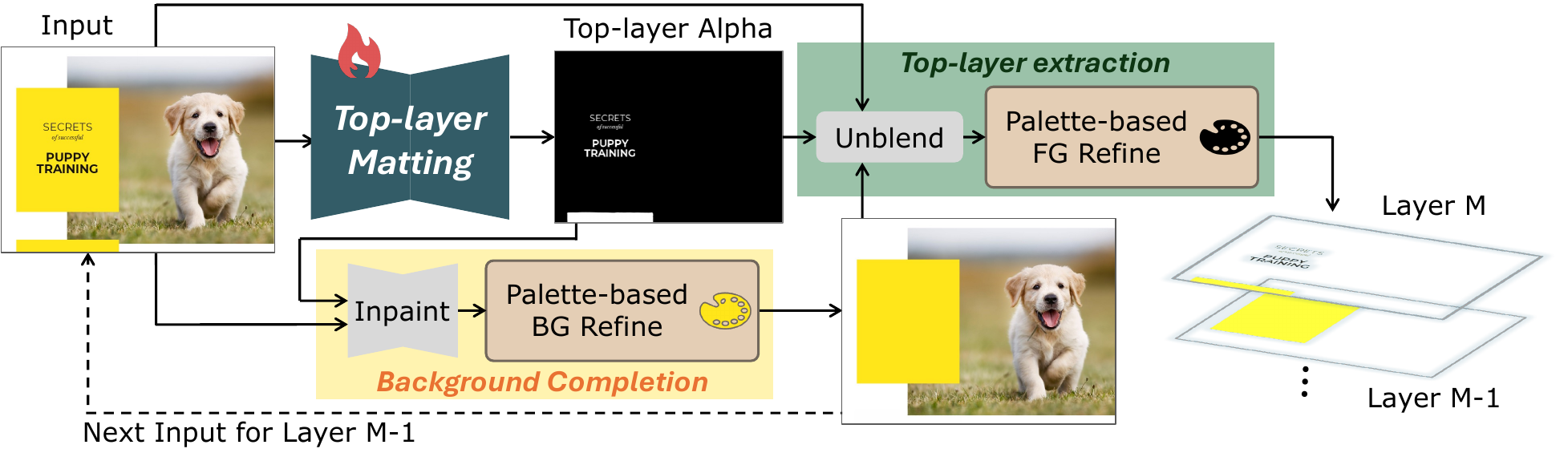}
    \caption{\ours{} decopmoses raster graphic designs into layers by iteratively extracting the top-layer and completing the background.
    Our training target is the top-layer matting model.
    \cref{fig:bg-refine,fig:fg-refine} illustrate details of the top-layer extraction and background completion.}
    \label{fig:pipeline}
\end{figure*}

\subsection{Image Layer Decomposition}
Image layer decomposition is a task to decompose an image into a sequence of layers, which are composable with a specific compositing function (e.g., alpha compositing) to reproduce or approximate the original image~\cite{porter1984compositing}.
Color segmentation represents an image with semi-transparent color layers, targeting digital paintings~\cite{tan2016decomposing} and natural images~\cite{tan2018efficient,aksoy2017unmixing,akimoto2020fast}.
Koyama \etal~\cite{koyama2018decomposing} propose to handle non-linear color blending functions, followed by the efficient deep learning-based extension~\cite{horita2022fast}.

There have been many studies on decomposing natural scenes at the object level~\cite{isola2013scene,monnier2021unsupervised,zhan2020self,zheng2021visiting,mulan,liu2024object}. For instance, PCNet~\cite{zhan2020self} decomposes a scene image into object layers by estimating the order of objects and the RGB of occluded parts.
While PCNet assumes the object modal mask is given, Zhang \etal~\cite{zheng2021visiting} create layered data including occluded parts in indoor scenes and decompose the image by training instance segmentation, depth estimation, and background completion.
Text2Layer~\cite{zhang2023text2layer} extracts salient objects from natural images using matting and generates training data for layered image generation.
\TODO{違いを明確に。ただしpreprint}
Recently, MULAN~\cite{mulan} decomposes natural images, including outdoor scenes where obtaining ground truth data is difficult, by combining the latest off-the-shelf open vocabulary object detection models~\cite{yao2023detclipv2}, zero-shot segmentation~\cite{kirillov2023segment}, depth estimation~\cite{ranftl2020towards}, and instance ordering~\cite{lee2022instance} with heuristics.
While the above studies mainly focus on object decomposition, Yang \etal~\cite{yang2024generative} decompose physical object effects (\eg, shadows or reflections) as well.

Compared to the natural image decomposition, graphic design decomposition has to deal with different granularities of \emph{objects}; \eg, a corporate logo in a graphic design consists of an illustration and a text, and whether they should be decomposed into parts depends on the context.
Considering the nature of the task, we propose a simple and effective method and a new quantitative evaluation protocol for inconsistent ground-truth.
A concurrent work~\cite{accordion} tackles the same task as ours with a stacked pipeline approach using a VLM trained on closed data.
\ours{}'s pipeline is overwhelmingly simple and leverages domain knowledge to refine the final quality.
We compare \ours{} with a VLM-based pipeline in our experiment.

\subsection{Image Vectorization}
Related to layer decomposition, image vectorization converts an image or a part into a set of parameters of a specific drawing function, rather than layer images.
Our layer decomposition approach can be useful for vectorization as a pre-processing step to extract part-based raster images.
Du \etal~\cite{du2023image} and Favreau \etal~\cite{favreau2017photo2clipart} obtain a sequence of linear gradient layers that approximate the original image by optimization using alpha blending.
Several works attempt to generate SVG-based representation from raster images~\cite{shen2021clipgen,ma2022towards,song2023clipvg,carlier2020deepsvg,reddy2021im2vec,rodriguez2023starvector},
where they typically assume vector art, cleanly masked images, or clean segmented images as input.
A few specifically focus on typographic representation in graphic design, where they estimate text rendering parameters~\cite{accordion,shimoda2021rendering}.

\subsection{Image Matting and Foreground Extraction}
\TODO{segmentationも?}
Image matting is a task to estimate alpha mattes of objects in an image, and together with other tasks such as background inpainting, forms the layer decomposition task.
Matting approach often assumes user-specified trimap~\cite{chuang2001bayesian,sun2004poisson,xu2017deep,yao2024vitmatte}, and a few trimap-free methods have been reported recently~\cite{birefnet,li2023matting}.
\ours{} mainly uses network architectures used in matting~\cite{birefnet} to extract unoccluded top layers.

While matting estimates alpha mattes, foreground color estimation involves determining the color of the foreground that is mixed with the background.
There are energy-based methods~\cite{levin2007closed,chen2013knn,aksoy2017designing} and their efficient versions~\cite{germer2021fast,forte2021approximate}, as well as deep learning-based methods~\cite{Lutz2021foreground} that estimate the foreground color given the alpha.
Hou \etal and Li \etal simultaneously estimate the alpha map and foreground color given an image and a trimap~\cite{hou2019context,li2025drip}.
The foreground color is deterministic when the background color and foreground alpha are given.
In our setup, we obtain the foreground matte from our trained model and the background color from high-quality background inpainting~\cite{lama}, and then calculate the foreground color.

    \section{Problem Formulation}

Graphic layer decomposition is the task of decomposing a raster graphic design image $\bm{x} \in [0, 1]^{H \times W \times 3}$ into a sequence of layers $Y=(\bm{l}_k \in [0,1]^{H \times W \times 4})_{k=0}^{K}$.
Here, $H$ and $W$ represent the height and width of the image, respectively. $\bm{x}$ is an RGB image, and $\bm{l}_k$ is an RGBA image, with 3 and 4 channels, respectively.
$k$ represents the blending order of the layer, \ie, the z-index.
$\bm{l}_{k>0}$ is the foreground layer, and $\bm{l}_0$ is the background layer.

The layer sequence $Y$ is composited by the following recursive process from $k=1$ to $k=K$ ($\bm{x} = \bm{x}_{K}$):
\begin{align}
    \label{eq:rasterize}
    \bm{x}^{\text{C}}_{k} &= {\rm B}(\bm{l}_{k}, \bm{x}^{\text{C}}_{k-1}),
    \\ &= \bm{l}^{\text{C}}_{k} \odot \bm{l}^{\text{A}}_{k} + \bm{x}^{\text{C}}_{k-1} \odot (1 - \bm{l}^{\text{A}}_{k}).
\end{align}
Here, the superscript $\text{A}$ represents the alpha channel, and $\text{C}$ represents one of the RGB channels. ${\rm B}(\cdot)$ is the alpha blending function, $\odot$ is element-wise multiplication, and $\bm{x}_{k}$ is the $k$-th blended image.

In this study, we solve the inverse problem of the above, \ie, layer decomposition that estimates the layer sequence $Y$ from the raster image $\bm{x}$.
The granularity of the layer depends on the dataset, and in this study, we treat the human-made graphic designs in the dataset as ground truth.

    \section{Approach}
\label{sec:method}

\ours{} solves the decomposition task by iterative extraction of \emph{top-layers}, which are not occluded by any other layers, and background completion (\cref{fig:pipeline}).
Our formulation integrates the subtasks of layer decomposition, which prior methods~\cite{mulan,accordion} separately address, into a single task, leading to a simplified implementation and performance gain by the simple training goal.
Additionally, we refine the final decomposition quality by leveraging the domain prior that graphic designs often contain simple, texture-less elements or backgrounds.

\subsection{Iterative Decomposition}
\label{sec:method:framework}

We obtain layer predictions $\hat{Y} = (\hat{\bm{l}}_m \in [0,1]^{H \times W \times 3} )_{m=0}^M$ from an input image $\bm{x}$ by iterative processes from front ($m=M$) to back ($m=1$) as follows:
\begin{align}
    \hat{\bm{l}}^{\text{A}}_{m} &= F_{\theta}(\hat{\bm{x}}_m) \\
    \hat{\bm{x}}_{m-1} &= G_{\phi}(\hat{\bm{x}}_m, \hat{\bm{l}}^{\text{A}}_{m}) \\
    \hat{\bm{l}}^{\text{C}}_{m} &= {\rm B^{-1}}(\hat{\bm{x}}^{\text{C}}_{m-1}, \hat{\bm{x}}^{\text{C}}_{m}, \hat{\bm{l}}^{\text{A}}_{m}) 
    \\ &= (\hat{\bm{x}}^{\text{C}}_{m} - \hat{\bm{x}}^{\text{C}}_{m-1} \odot (1 - \hat{\bm{l}}^{\text{A}}_{m})) \oslash \hat{\bm{l}}^{\text{A}}_{m} \label{eq:unblend}
\end{align}
where $\hat{\bm{x}}_M = \bm{x}$, $\hat{\bm{l}}_0 = \hat{\bm{x}}_0$, and $\oslash$ is an element-wise division. The superscripts $\text{A}$ and $\text{C}$ denote the alpha channel and one of the RGB channels, respectively.
$F_{\theta}(\cdot)$ is a model that takes an image as input and outputs an alpha map, which is the same as the trimap-free matting task, as long as the matting target is the top-layers.
The output alpha contains all top layers; they are decomposed in one iteration.
$G_{\phi}(\cdot)$ is a background completion model that takes an image and a target mask obtained from the top-layers alpha map as input and outputs an image with the target area completed.
The background completion model should not insert new objects.
We tried several inpainting approaches, including generative model-based completion~\cite{blackforest2024fluxfill}, and found that generative approaches often insert unnecessary objects.
We use LaMa~\cite{lama} for $G_{\phi}$, which satisfies our inpainting requirement.
$\rm B^{-1}(\cdot)$ is a process that estimates the RGB values from the completed background and the alpha map of top-layers.
Since we know the alpha map and the completed background, we can calculate the RGB values of top-layers by simple arithmetic as the inverse process of alpha blending $B(\cdot)$.
Note that existing methods~\cite{mulan,accordion} replace the alpha of the original image with the predicted soft or hard segmentation mask.
For pixels with an alpha of 1, our method estimates the same RGB values as these naive methods.
However, for other transparent pixels, our method estimates the RGB values while considering blending with the background.
This primarily improves the quality of layer boundaries where soft blending is applied.
We terminate the iteration (\ie, $m=M$) when there are no pixels above a certain threshold in the matting result $\hat{\bm{l}}^{\text{A}}_m$.

\subsection{Training}
\label{sec:method:training}
In \ours{}, we utilize two learnable models: top-layer matting model $F_{\theta}$ and background inpainting model $G_{\phi}(\cdot)$.
Since image inpainting is a general task and reasonably performant models are available in the public~\cite{lama}, we use an off-the-shelf pretrained model without fine-tuning.
For training the top-layer matting model, we prepare pairs of an input RGB image $\bm{x}$ and a target alpha map of top-layers $\bm{l}^{\text{A}}$ from Crello~\cite{yamaguchi2021canvasvae} for supervised learning.
We first check the occlusion of each layer based on the layer information and integrate the alpha maps of non-occluded layers into a single alpha map.
This clear target definition eliminates ambiguity in training.
Similarly to \ours{}'s pipeline, we create multiple pairs per design sample by recursively performing the same process on the remaining background.

We follow the prior study~\cite{birefnet} and define the loss function as below:
\begin{align}
    \mathcal{L}(\hat{\bm{l}}^{\text{A}}, \bm{l}^{\text{A}}) &= \lambda_{\text{BCE}} \mathcal{L}_{\text{BCE}}(\hat{\bm{l}}^{\text{A}}, \bm{l}^{\text{A}}) \notag \\ 
    & + \lambda_{\text{IoU}}\mathcal{L}_{\text{IoU}}(\hat{\bm{l}}^{\text{A}}, \bm{l}^{\text{A}}) + \lambda_{\text{SSIM}} \mathcal{L}_{\text{SSIM}}(\hat{\bm{l}}^{\text{A}}, \bm{l}^{\text{A}}), 
\end{align}
where $\mathcal{L}_{\text{BCE}}(\cdot)$, $\mathcal{L}_{\text{IoU}}(\cdot)$, and $\mathcal{L}_{\text{SSIM}}(\cdot)$ are binary cross-entropy, Intersection-over-Union (IoU) loss, and structural similarity index (SSIM) loss, respectively, and
$\lambda_{\text{BCE}}$, $\lambda_{\text{IoU}}$, and $\lambda_{\text{SSIM}}$ are weights for each loss term.
We train the matting model using all loss functions at the early steps and then use only the SSIM loss to improve the boundary quality.

During inference, the model takes the background completion result ($\hat{\bm{x}}_m$) as input instead of the clean intermediate composite result, except for the first iteration.
This gap between training and inference can degrade the decomposition quality.
To make the matting model robust to the inpainting artifacts, we include training examples where the top-layer regions are completed by the background completion model.
Since the completion in areas spanning the back layers can alter their shape, leading to inconsistencies with the ground truth, we do not complete such areas when making training data.

\subsection{Palette-based Refinement}
Graphic designs often contain flat elements or backgrounds with few textures, such as decorations, texts, and vector shapes.
Based on this observation, we introduce a simple refinement approach that greatly improves the resulting appearance at the end of each iteration.

\begin{figure}[t]
    \centering
    \includegraphics[keepaspectratio, width=1\linewidth]{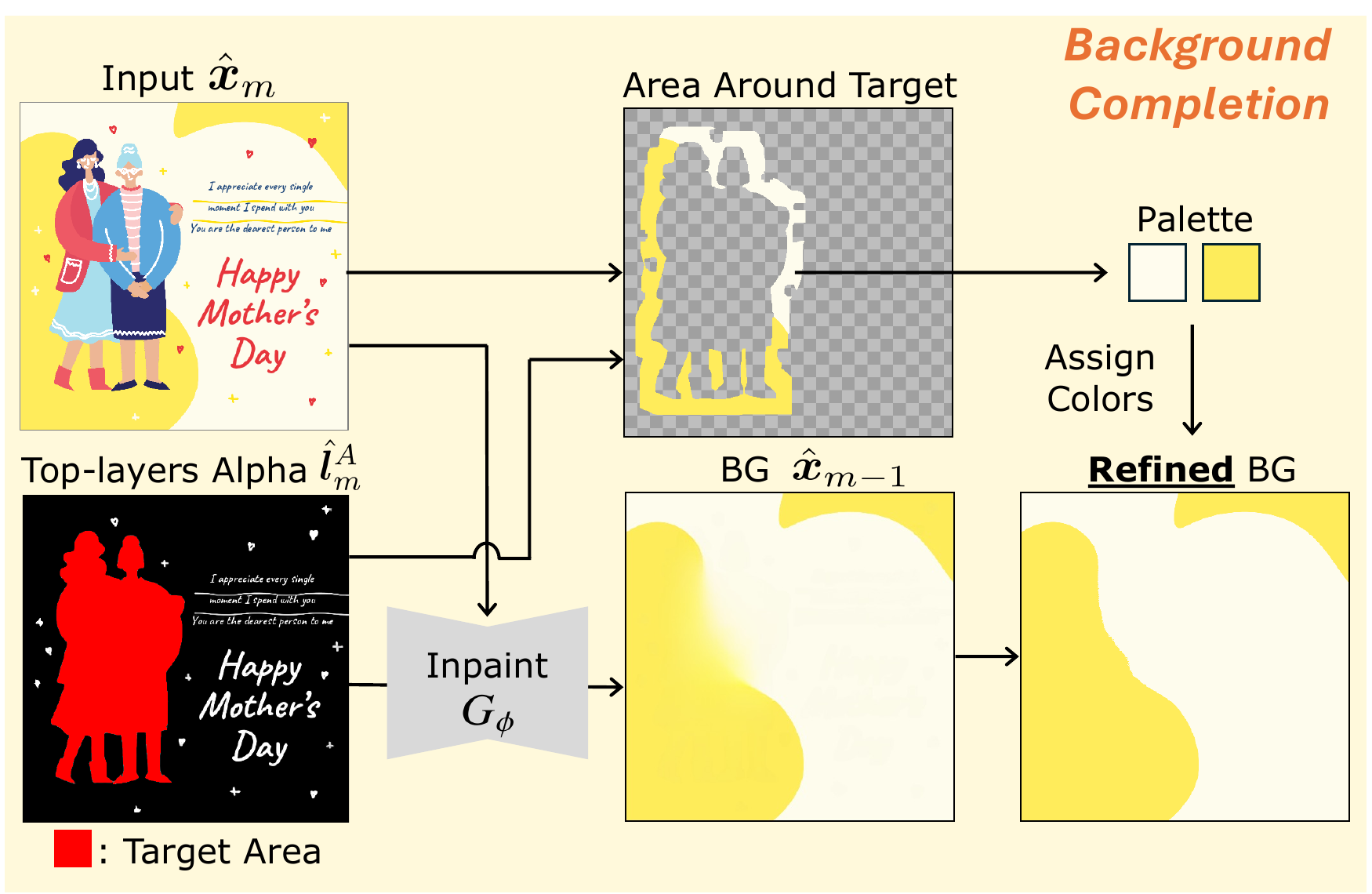}
    \caption{
        Background completion with palette-based refinement.
        We first complete the area of the predicted alpha map, then refine the target connected region based on the color palette of the surrounding area.
        We select the target area based on the color gradient of the surrounding area (shown in red).
        }
    \label{fig:bg-refine}
\end{figure}

\paragraph{Background refinement (\cref{fig:bg-refine})}
We first divide the alpha map into connected regions and process each area.
We calculate the color gradients of the surrounding area of the completion target area.
If the area with zero color gradients is dominant, we assume that the completion target region is a flat-paint area with few textures.
We extract the dominant colors, \ie, the palette, of the region based on percentiles and assign the completed RGB values to the nearest palette color in the Lab color space.
The background completion model can make rough predictions for such flat backgrounds even with noticeable artifacts, allowing our simple refinement to work effectively.
Our refinement eliminates such artifacts.

\begin{figure}[t]
    \centering
    \includegraphics[keepaspectratio, width=1\linewidth]{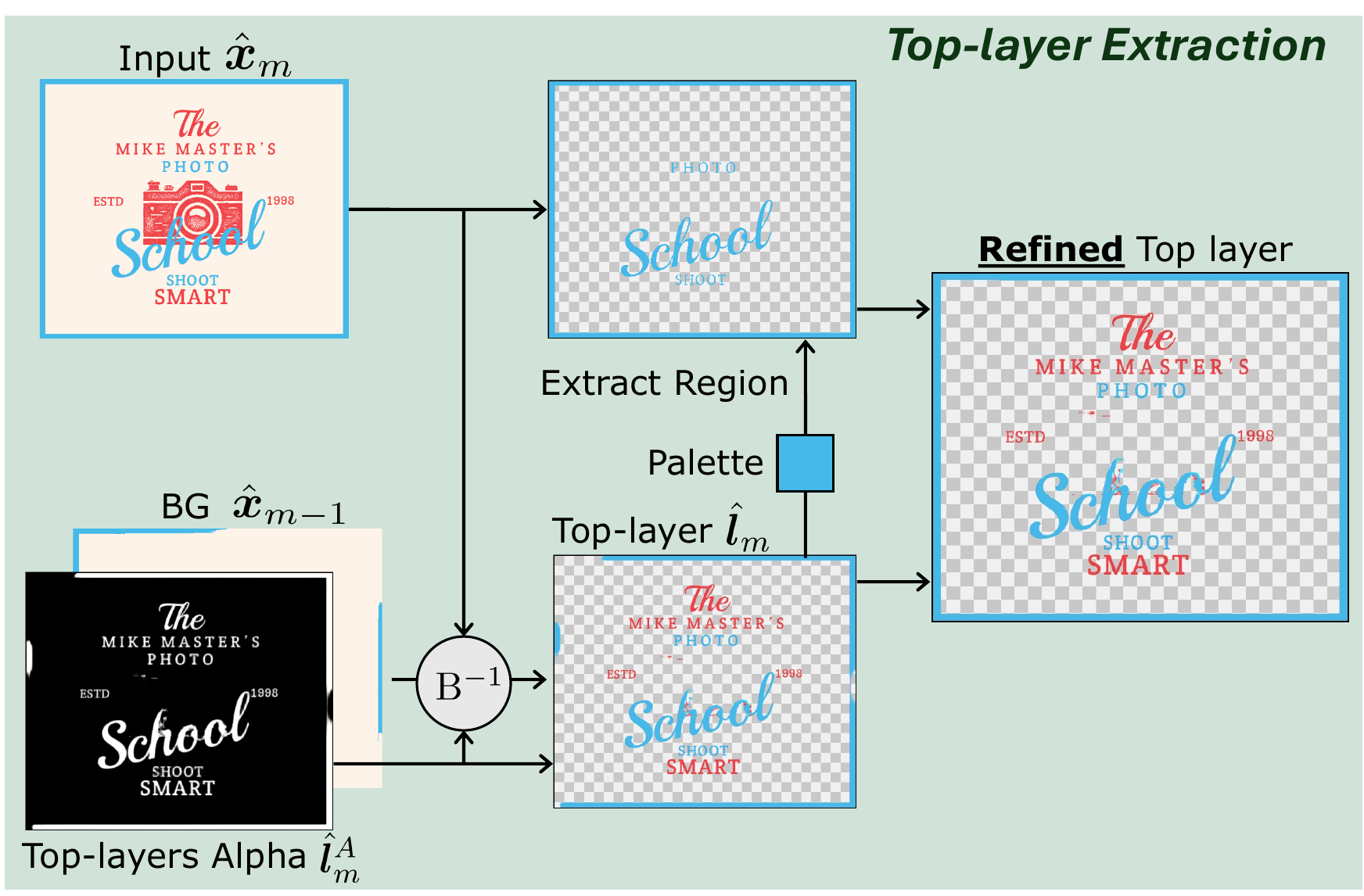}
    \caption{
        Palette-based foreground refinement. 
        First, we estimate the RGB values of the top-layer using the input image, the top-layer's alpha, and the background by the unblend function.
        Next, we extract the color palette of the connected components of the top-layer, extract the region that matches the color from the original image, integrate the connected color region with a large overlap with the predicted alpha map, and use it as a new alpha map.
        Note that the missing blue edge is refined in this figure.
    }
    \label{fig:fg-refine}
\end{figure}

\paragraph{Foreground refinement (\cref{fig:fg-refine})}
Similar to background refinement, we first divide the alpha map into connected regions.
Next, we calculate the color gradients within each region and classify regions where the area with zero color gradients dominates as flat regions.
We extract the region that matches the palette color from the input image ($\bm{x}$) or intermediate completed background ($\hat{\bm{x}}_m$) and select the region if the overlap with the predicted alpha exceeds a threshold.
Then, we integrate the selected regions as a new mask.
Since the obtained mask is binary, we calculate the alpha value around the mask using the palette color and the background color to derive the final alpha value.
We often observe that this refinement improves the quality of the boundary parts and thin decoration layers (\eg, lines and frames), which the plain matting model frequently fails.

    \section{Decomposition Metrics}
\label{sec:metrics}
There are two problems in evaluating the quality of predicted layers $\hat{Y}$ against the ground truth $Y$.
First, the number of layers in the ground truth and the predicted layers can differ, making it non-trivial to compare directly.
To address this, we apply order-aware layer alignment using DTW~\citep{Mueller07_InformationRetrieval_SPRINGER}.
Second, the quality of the predicted layers can be evaluated from two perspectives: visual quality and granularity.
If we do not consider these aspects separately, we may underestimate the quality of the predicted layers due to differences in granularity, even if they are practically useful. 
We measure granularity by the number of edits required to align the two; we allow merging adjacent layers in the z-index and report both the number of edits and the visual quality after the editing operations.

\paragraph{Layer alignment}
\label{sec:layer_alignment}
As pre-processing, we first group the ground truth and predicted layers based on visibility.
Specifically, we extract layers whose visible regions (\ie, alpha values greater than zero) are not occluded by any other higher layers in z-index, blend them into a single layer, and repeat the same operation with the remaining layers.
This operation never affects the appearance of the composite image and forms what we refer to as a top-layer.

Next, we find alignment between the two layer sequences with different lengths using DTW, which considers the sequence order even if the lengths differ.
We obtain a set of pairs $P=\{(k_s,q_s)|s=0,1,\ldots,S\}$, where $k_s$ and $q_s$ represent the layer indices, and $S$ is the number of pairs.
Note that resolved pairs satisfy the monotonicity condition, \ie., $k_s$ and $q_s$ are increasing sequences; in other words, layers cannot be shuffled during alignment (see \suppref{sup:dtw}).
We define the distance metric for the layer pair as the sum of the negative value of the alpha's soft IoU and 
L1 distance of the RGB channels weighted by the ground-truth alpha, as introduced in \cite{suzuki2024fast}.

Finally, we compute the quality metric between the two layer sequences as follows:
\begin{align}
    \label{eq:metric}
    \mathcal{E}(\hat{Y}, Y) = \frac{1}{S}\sum_{s=0}^{S} e(\hat{\bm{l}}_{k_s}, \bm{l}_{q_s}),
\end{align}
where $e(\cdot)$ is an arbitrary function that measures the similarity or distance between layers. 
We use the weighted L1 distance of the RGB channels and the soft IoU of the alpha channel as $e(\cdot)$, similar to DTW's distance metric.

\paragraph{Layer merge}

Due to the ill-posed nature of layer decomposition, decomposition results sometimes do not align well with the ground truth. \TODO{Provide examples?}
In this work, we relax the alignment constraints by allowing \emph{edits}.
The idea is inspired by minimum edit distance~\citep{wagner1974string}, which is commonly used for string alignment. We define a specific edit operation set for layers, and report both the maximum number of allowed edits and the distance metric used in DTW after edits.
This gives a straightforward insight into how many layer-level edits are required for good alignment.

For simplicity, we define a single edit operation; \texttt{Merge}, which merges two consecutive layers in z-index, when the edit yields the highest positive distance improvement.
We apply edits iteratively until no further improvements are possible or the number of layers is reduced to 2.
The ground truth is also mergeable.
Visual examples of the edit process can be found in \suppref{fig:sup-edit-process-1,fig:sup-edit-process-2}.

    \section{Experiments}
\label{sec:experiments}

\begin{figure*}[t]
    \centering
    \begin{subfigure}{0.495\linewidth}
        \centering
        \includegraphics[width=\linewidth]{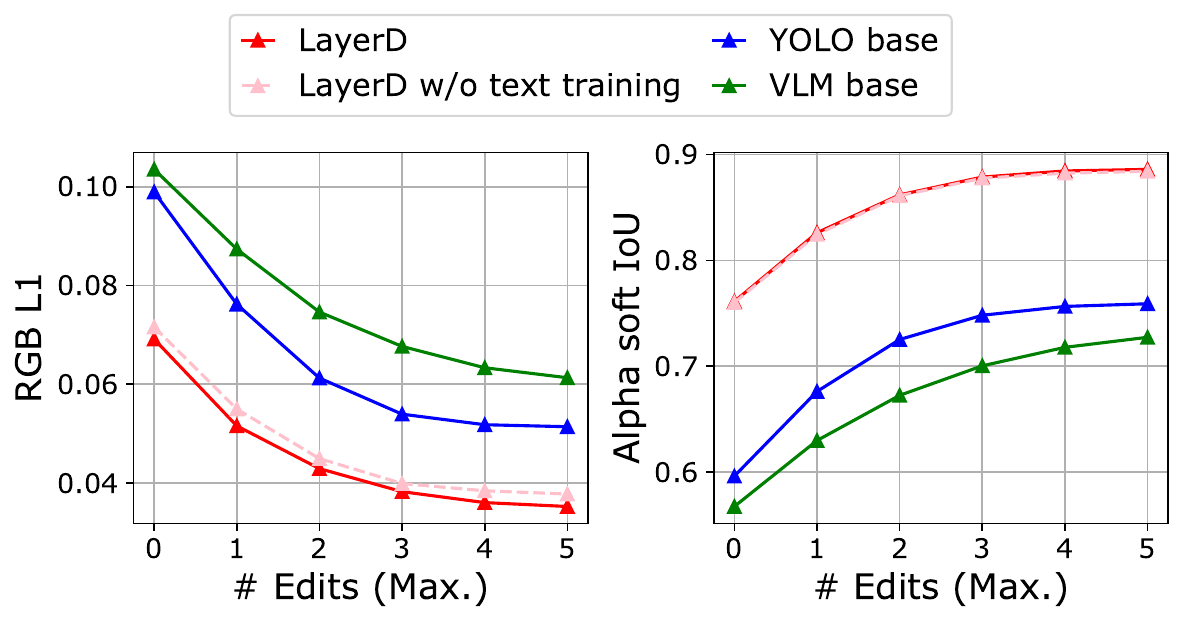}
        \caption{Without text layers}
        \label{fig:main_comparison_wo_text}
    \end{subfigure}
    \hfill
    \begin{subfigure}{0.495\linewidth}
        \centering
        \includegraphics[width=\linewidth]{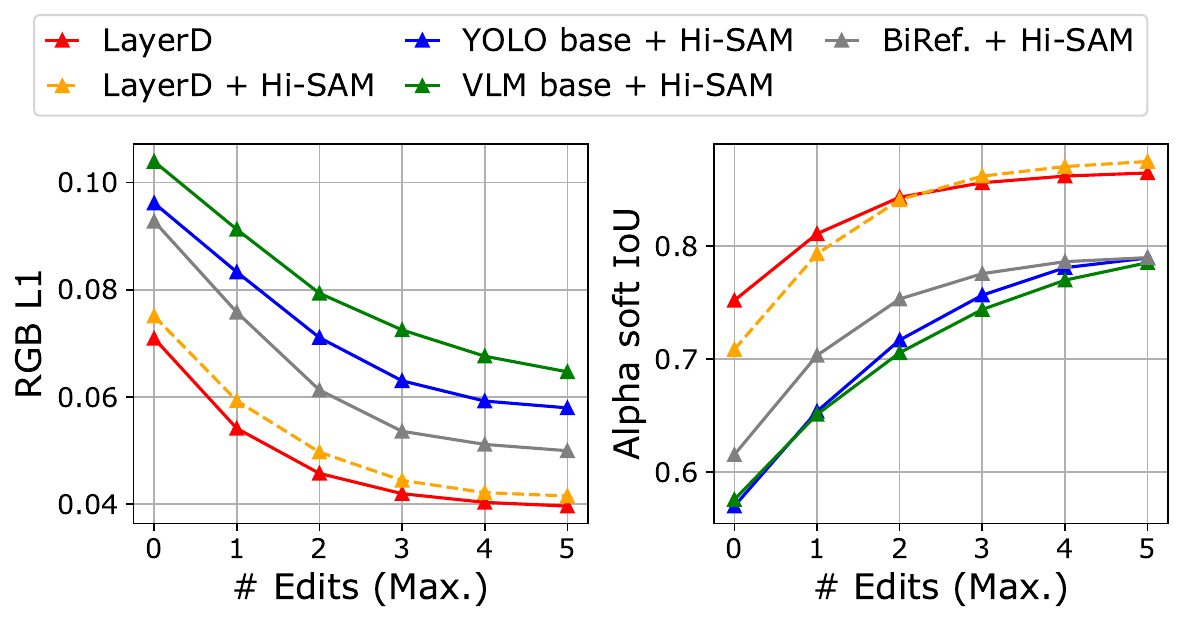}
        \caption{All layers}
        \label{fig:main_comparison_all}
    \end{subfigure}
    \caption{Baseline comparisons. We show visual quality metrics (RGB L1, Alpha IoU) as the maximum number of allowed edits increases. The left two are the results when we exclude text layers from the dataset, and the right two are the results when all layers are included. ``w/o text training'' indicates the case where text layers are not included during training.}
    \label{fig:main_comparison}
\end{figure*}

\subsection{Datasets}
We use the Crello dataset~\cite{yamaguchi2021canvasvae}, which is a collection of graphic design templates, for both training and evaluation.
We obtain pairs of input images and their ground-truth layer sequences from the layer structure in design templates.
We follow the data split of v5.1.0 %
to obtain 19,478 / 1,852 / 1,971 samples for train, validation, and test split, respectively.
We resize all images to maintain their aspect ratio, with the shorter side set to 512.
In this work, we exclude transparent layers from the evaluation since neither our method nor the baselines primarily focuses on accurate transparency estimation. %
For all methods, we conduct training and validation on the train and validation split, respectively, and report the results on the test split.
For training of \ours{} as described in \cref{sec:method:training}, we generate input / ground-truth pairs, and finally obtain 48,725 and 4,674 pairs for the train and validation, respectively.
As typography is one of the unique domain properties, we prepare the full dataset and the variant that excludes all text layers for evaluation.

\subsection{Baselines}
Although there are a few methods comparable to \ours{}, none of them have publicly available code or models.
We design the following baseline and implement an existing approach with minor modifications to fit our setting.
Additionally, we evaluate all methods using Hi-SAM~\cite{ye2024hi}, a state-of-the-art text segmentation, for initial layer extraction.

\paragraph{YOLO baseline}
We design a naive baseline that combines state-of-the-art object detection and pretrained segmentation models.
First, since most graphic designs contain text on the top, we segment text using Hi-SAM~\cite{ye2024hi} and complete the background using LaMa~\cite{lama}.
Next, we detect bounding boxes of layers from the remaining background.
To this end, we extract bounding boxes from the layer structure of Crello and fine-tune YOLO~\cite{wang2024yolov9} using them.
We determine the z-index of layers based on heuristics in graphic design, assuming that the smaller box is in front when the boxes overlap.
Then, we obtain the segmentation masks of the topmost boxes using a pretrained SAM2~\cite{ravi2024sam} and perform background completion.
We repeat this process, except for text segmentation, until the number of detections becomes zero.
We obtain the final layers by replacing the alpha of the input or completion image with the predicted segmentation mask and blacking out the color of pixels with an alpha close to zero.

\paragraph{VLM baseline}
We also consider a baseline that follows the approach of Accordion~\cite{accordion}.
Accordion generates layered graphic design by combining raster-based image generation~\cite{flux} and layer decomposition, where VLM first takes an image as input and generates a decomposition plan with a JSON-like description of bounding boxes and z-indices of layers, and then applies segmentation and background completion in a front-to-back order to obtain the layer sequence.
Since the model and training data are not publicly available, we reproduce this with minor modifications in our experiment.
We use PaliGemma2~\cite{steiner2024paligemma} as the backbone VLM and fine-tune on the Crello train set, Hi-SAM for text detection, and SAM2 for other elements.
For background completion, we use LaMa~\cite{lama} to ensure fairness with other methods.

\subsection{Implementation Details}
We use BiRefNet~\cite{birefnet} with Swin-L~\cite{liu2021swin} pre-trained on natural image object segmentation 
for the top-layers matting model, and train it on Crello for 60 epochs with a batch size of 12.
We set the maximum number of iterations for decomposition to 3 for \ours{} and the YOLO baseline.
The maximum number of colors for palette-based refinement is set to 10 for the foreground and 2 for the background.
For evaluation, we change the maximum number of allowed edits from 0 to 5, and report the visual quality for each case.

\begin{figure}[t]
    \centering
    \includegraphics[keepaspectratio, width=1.0\linewidth]{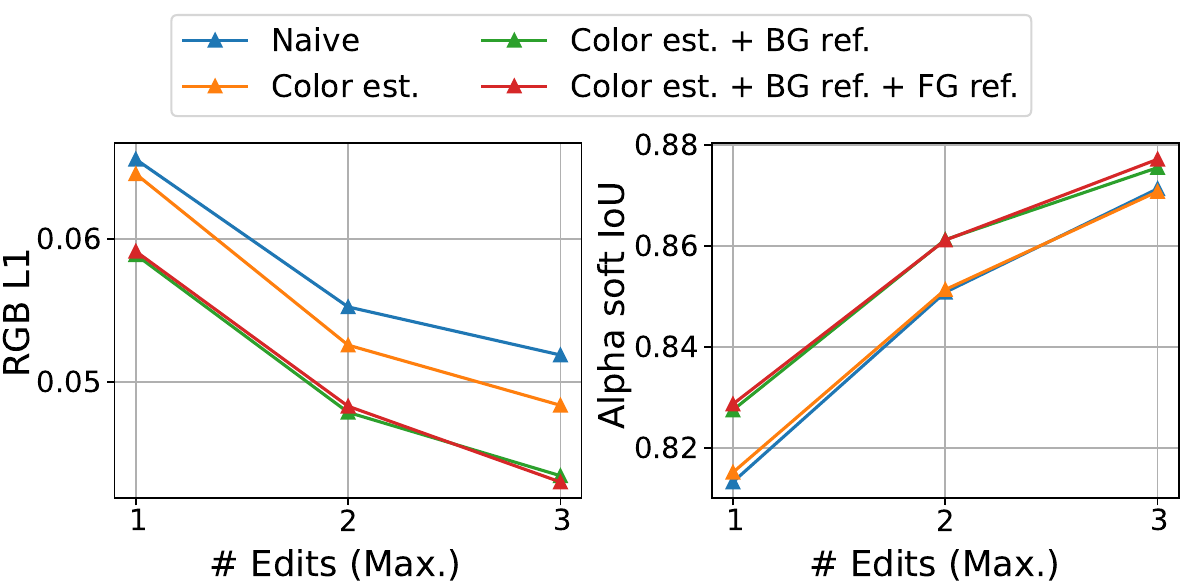}
    \caption{Ablation results of foreground color estimation and refinement.
    }
    \label{fig:refine_ablation}
\end{figure}

\begin{figure}[tb]
    \centering
        \includegraphics[width=\linewidth]{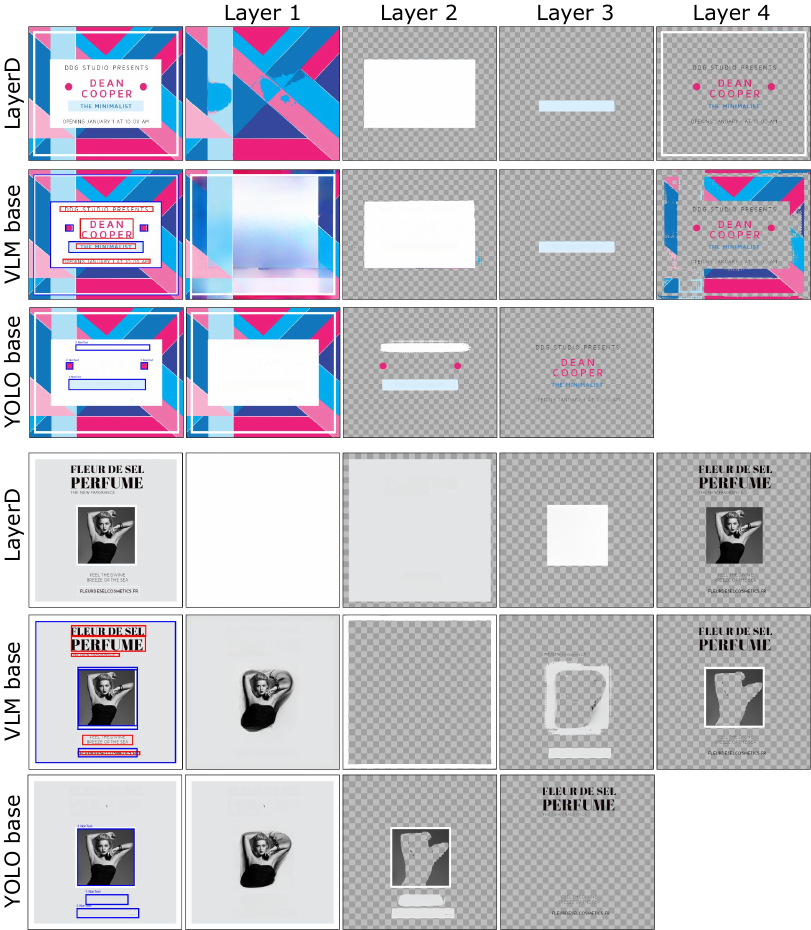}
    \caption{Comparison of decomposition results by \ours{} and baselines. 
    The leftmost images are the input image (\ours{}), object detections (VLM), and text-removed input (YOLO).
    Red rectangles indicate text, and blue rectangles indicate other elements.
    }
    \label{fig:vis_main_comparison}
\end{figure}

\subsection{Quantitative Evaluation}

\paragraph{Baseline comparison}
We compare \ours{} with baselines with and without text layers in \cref{fig:main_comparison_all} and \cref{fig:main_comparison_wo_text}, respectively.
In all metrics, \ours{} generates layer sequences close to the ground truth with fewer edits.
Our simplified pipeline and training objective are effective in layer decomposition.
Moreover, in the results for all layers (\cref{fig:main_comparison_all}), \ours{} alone shows higher performance than \ours{} + Hi-SAM, which replaces the first iteration with Hi-SAM.
\ours{}, which is specifically trained for graphic design, is more effective than Hi-SAM, which is trained for text segmentation without being limited to graphic design.
More interestingly, in the case of decomposition without text layers, as shown in \cref{fig:main_comparison_wo_text}, \ours{} trained with text layers exhibits slightly better performance than \ours{} trained without text, even though the decomposition targets contain no text.
We suspect that text is essentially a variant of vector shapes, and training with text layers improves the decomposition performance of these elements.
Our method outperforms the BiRefNet without additional training in \cref{fig:main_comparison_all}, indicating the importance of training on top-layer matting.

\paragraph{Refinement ablation}
\cref{fig:refine_ablation} shows the ablation results of foreground color estimation and refinement.
First, the color estimation by the inverse blending (\cref{eq:unblend}) reduces the RGB L1 compared to the ``Naive'', which simply replaces the alpha with the predicted mask.
Background refinement significantly improves the RGB L1, resulting in better subsequent layer decomposition, as indicated by the improvement in Alpha IoU.
The foreground refinement slightly improves the quality of the alpha map. Although the quantitative improvement is slight, we observe that foreground refinement improves the quality of boundary regions.

\subsection{Qualitative Evaluation}

\paragraph{Baseline comparison}
\cref{fig:vis_main_comparison} shows the qualitative comparison of \ours{} with VLM and YOLO baseline.
The VLM baseline, which relies on bounding boxes, struggles with proper decomposition when detection fails or when the detected boxes overlap.
The bounding box of a layer with a large hole like a donut includes the elements of the entire image, making it difficult for the subsequent segmentation (\cref{fig:vis_main_comparison} top sample).
The YOLO baseline suffers from false negatives in detection, resulting in incomplete decomposition.
In contrast, \ours{} directly extracts layers without relying on bounding boxes, resulting in clean decomposition results in all cases.

\paragraph{Refinement effect}

\begin{figure}[tb]
    \centering
        \includegraphics[width=1\linewidth]{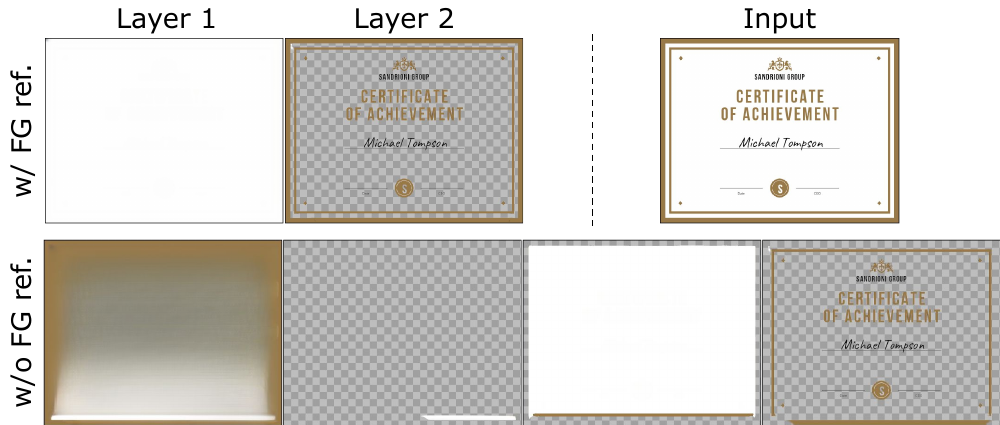}
    \caption{Example with or without foreground refinement. Without refinement, layers tend to have a collapsed boundary.}
    \label{fig:vis_fg_refine}
\end{figure}

\begin{figure}[tb]
    \centering
        \includegraphics[width=1\linewidth]{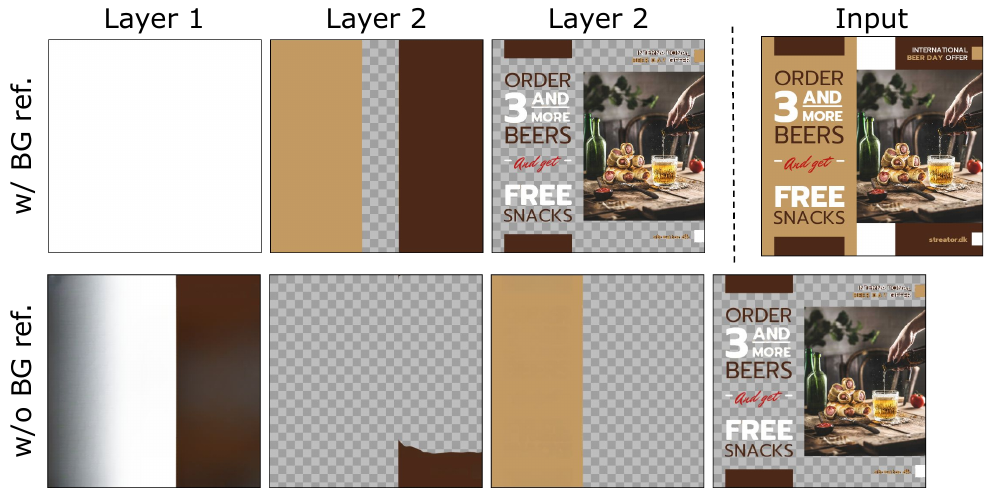}
        \caption{Example with or without background refinement. Our refinement prevents the background from being divided into segments or filled with unexpected colors.}
    \label{fig:vis_bg_refine}
\end{figure}

\cref{fig:vis_fg_refine} shows an example of foreground refinement.
The foreground refinement recovers the large missing gold decoration with a large hole, which is a typical failure case of the top-layers matting model.
In \cref{fig:vis_bg_refine}, we show an example of background refinement.
Background refinement successfully completes the missing part of the background.
Since layer decomposition is recursive, failures in the previous iteration have a negative impact on subsequent iterations.
Our refinement contributes not only to the quality of the target layer itself but also to the quality of the subsequent layers, \ie, the overall quality of the layer decomposition.

\begin{figure}[tb]
    \centering
    \includegraphics[keepaspectratio, width=1\linewidth]{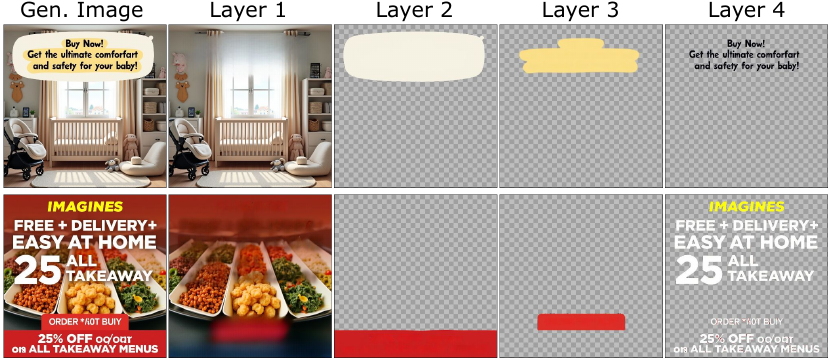}
    \caption{Decomposition results of \ours{} on raster images generated by FLUX.1 [dev]~\cite{flux}.}
    \label{fig:genai}
\end{figure}

\subsection{Applications}

In this section, we demonstrate that \ours{} enables a few applications out of the box.

\paragraph{Decomposing generated images}
\cref{fig:genai} shows the decomposition results of \ours{} on graphic design generated by FLUX.1 [dev]~\cite{flux}, which is one of the state-of-the-art text-to-image generator, using prompts from DESIGNERINTENTION-v2 benchmark~\cite{jia2023cole}.
Note that quantitative evaluation is not possible in this setup as ground truth layers are not available.
The results suggest that \ours{} can generalize and successfully decompose generated images, enabling an editable workflow for raster image generators.

\paragraph{Image editing}
We show the image editing examples using the decomposed layers by \ours{} in \cref{fig:teaser} and \suppref{sup:application}.
Here, we only perform simple operations such as color conversion, translation, and resizing at the layer level.
Although the layers obtained by \ours{} are grouped by top-layers, we can easily divide them into connected components for more granular editing.
Our decomposition enables flexible and intuitive layer-based editing.
Note also that there is no significant artifact in the editing results.

    \section{Conclusion}

We present \ours{} for decomposing raster graphic designs, where we propose the iterative extraction of unoccluded layers and background completion as well as refinement tailored for graphic materials.
We propose an evaluation protocol for the ill-posed decomposition task, where we introduce the notion of layer edit to quantify the difference from the unreliable ground truth.
The experiment showed that \ours{} led to solid improvement over baselines.

Our approach aims at decomposition, but it might be interesting to further consider vectorization~\cite{ma2022towards,song2023clipvg,carlier2020deepsvg,reddy2021im2vec}, which would expand the possible creative workflow.
In the other direction, our method could help learn a layered design generation model~\cite{jia2023cole,inoue2024opencole} as a pre-processing component, or be combined with recent automatic layered design editing~\cite{patnaik2025aesthetiqenhancinggraphiclayout,lin2024elementsdesignlayeredapproach}, or animation generation~\cite{liu2024logomotion} for interesting applications.

    { \small \bibliographystyle{ieeenat_fullname} \bibliography{main} }

    \clearpage
\setcounter{figure}{0}
\renewcommand{\thefigure}{\Alph{figure}}
\renewcommand{\thetable}{\Alph{table}}
\renewcommand{\thealgorithm}{\Alph{algorithm}}
\maketitlesupplementary
\appendix

\section{Editing Examples}
\label{sup:application}
\begin{figure*}[h]
    \centering
    \includegraphics[keepaspectratio, width=0.89\linewidth]{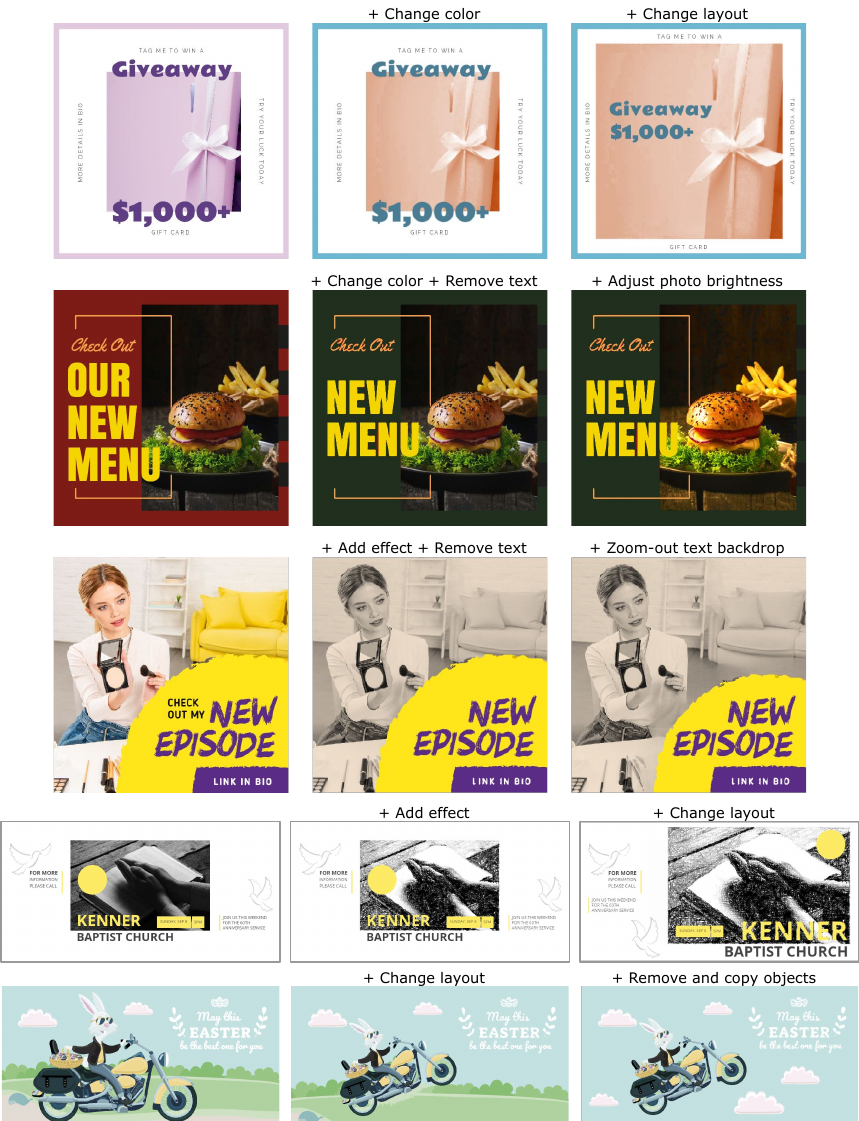}
    \caption{Editing examples on Crello~\cite{yamaguchi2021canvasvae} test set.
    The leftmost images are the original images, and the remaining images are edited ones based on the decomposed layers. We use \ours{} to decompose the original images into layers, divide them into connected components, and group text components using CRAFT~\cite{baek2019character}.
    Then, we perform various \emph{layer-level} edits, from simple layout changes to applying built-in image effects, on PowerPoint.}
    \label{fig:application-examples}
\end{figure*}

We show editing examples in \cref{fig:application-examples}.
Here, we use \ours{} to decompose the input image into layers, divide each layer into connected components, and group text components using CRAFT~\cite{baek2019character} to facilitate editing.
We import the layers into PowerPoint\footnote{\url{https://www.microsoft.com/powerpoint}} and perform various edits, from simple layout manipulation to applying built-in image effects, \emph{at the layer level}.
As the examples show, once the images are decomposed, users can intuitively edit them with precise control over each graphic element.

\section{Additional Results}
\label{sup:results}
\begin{figure*}[h]
    \centering
    \includegraphics[keepaspectratio, width=0.85\linewidth]{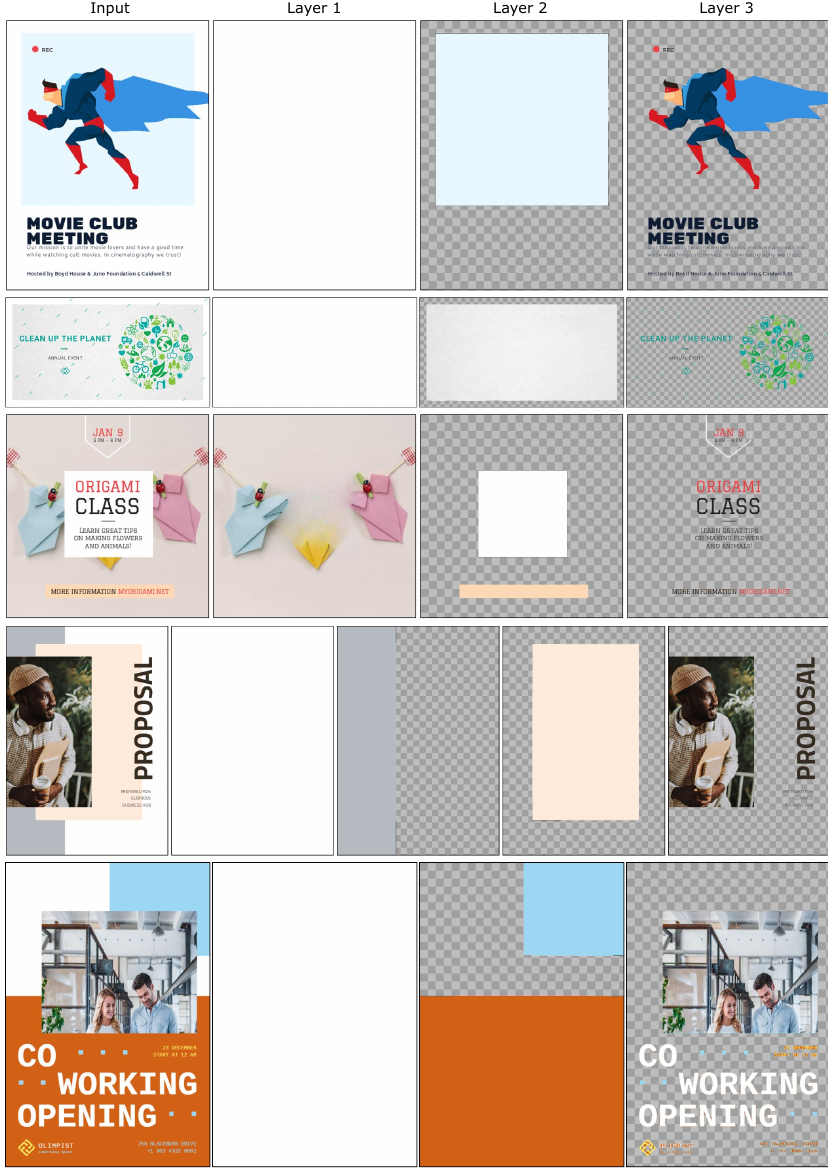}
    \caption{Additional qualitative results of our method on Crello~\cite{yamaguchi2021canvasvae} test set.
    The leftmost column shows the input image, and the remaining columns show the decomposed layers from back to front.
    }
    \label{fig:sup-qualitative-1}
\end{figure*}

\begin{figure*}[h]
    \centering
    \includegraphics[keepaspectratio, width=0.85\linewidth]{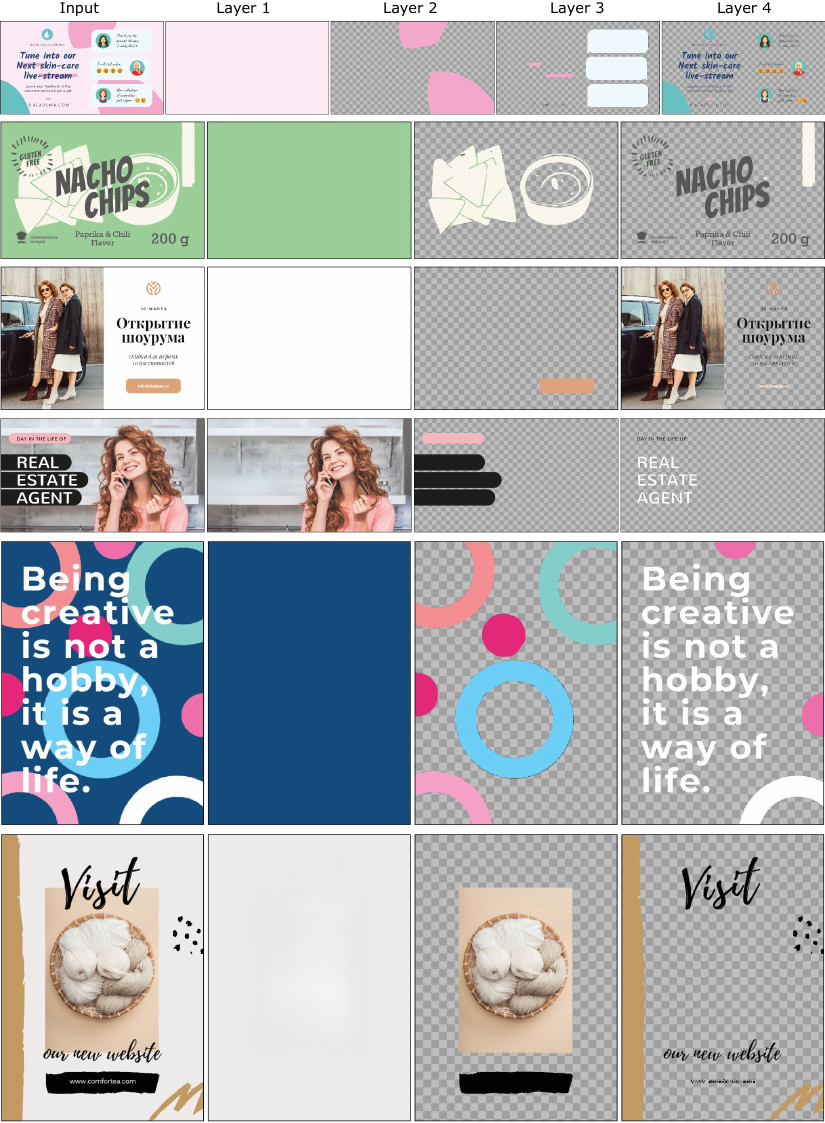}
    \caption{Additional qualitative results of our method on Crello~\cite{yamaguchi2021canvasvae} test set.
    The leftmost column shows the input image, and the remaining columns show the decomposed layers from back to front.
    }
    \label{fig:sup-qualitative-2}
\end{figure*}

We present additional examples of decomposed graphic design images using our method in \cref{fig:sup-qualitative-1,fig:sup-qualitative-2}.
These examples are selected from the Crello~\cite{yamaguchi2021canvasvae} test set and demonstrate the effectiveness of our method across diverse design styles.

\section{Failure Cases}
\label{sup:failure}
\begin{figure*}[h]
    \centering
    \includegraphics[keepaspectratio, width=0.82\linewidth]{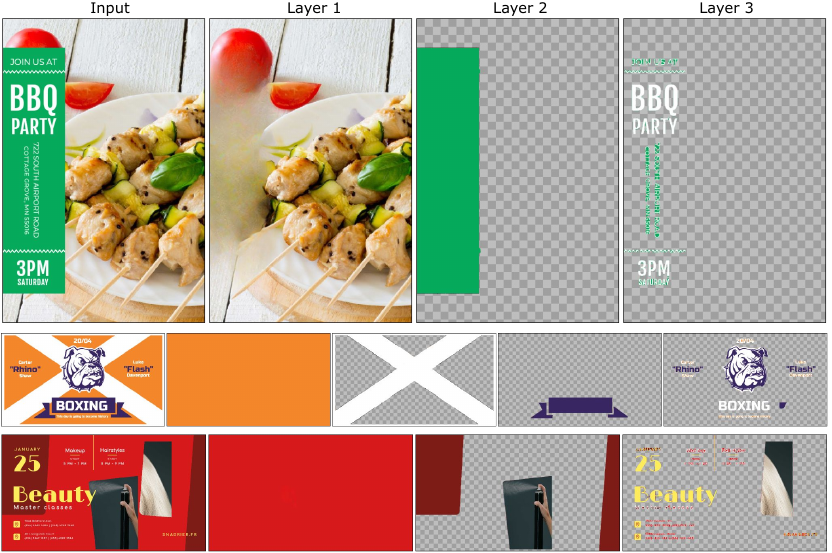}
    \caption{Failure samples for too small objects on Crello~\cite{yamaguchi2021canvasvae} test set.
    The leftmost column shows the input image, and the remaining columns show the decomposed layers from back to front.
    }
    \label{fig:sup-failure-1}
\end{figure*}

\begin{figure*}[h]
    \centering
    \includegraphics[keepaspectratio, width=0.82\linewidth]{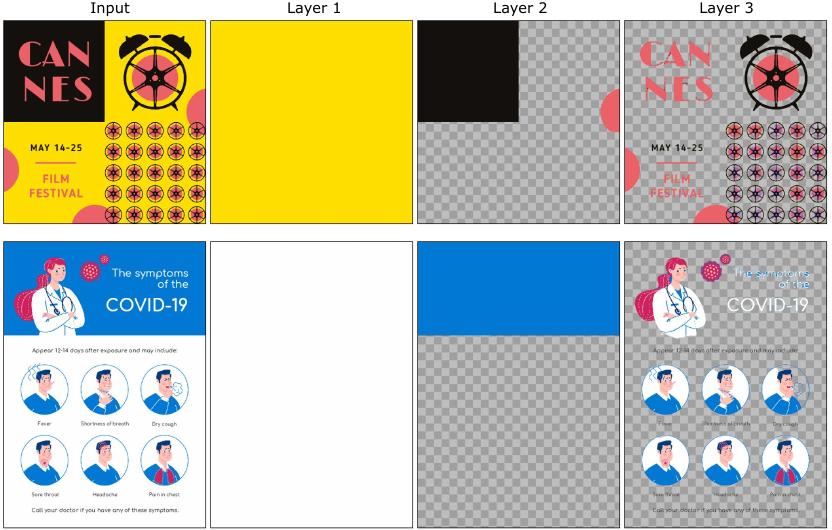}
    \caption{Failure samples due to the ambiguity of the layer granularity on Crello~\cite{yamaguchi2021canvasvae} test set.
    The leftmost column shows the input image, and the remaining columns show the decomposed layers from back to front.
    }
    \label{fig:sup-failure-2}
\end{figure*}

In \cref{fig:sup-failure-1,fig:sup-failure-2}, we show typical failure cases of our method.
The first set of failure cases (\cref{fig:sup-failure-1}) involves objects that are too small, such as detailed text descriptions, which are challenging to decompose due to their limited spatial extent.
We believe that these can be mitigated by increasing the resolution of the input images.
The second set of failure cases (\cref{fig:sup-failure-2}) is due to the ambiguity of the layer granularity.
For these samples, it is difficult even for humans to decompose them into the same layers consistently.
Although our evaluation metrics account for such ambiguity, we may need to improve training objectives or the post-refinement process to address these cases.

\section{User Study}
\label{sup:userstudy}
We conduct a user study in which 21 cloudworkers experienced in layer-based image editing rate the practical utility of 50 decomposition results---randomly ordered and anonymized---from \ours{} and our two baselines on the same images using a five-point scale.
\cref{tab:user-study} summarizes the results of the user study.
\ours{} achieves the highest average score, and a significant majority of the users (71.4\%) rate \ours{} the highest average score.
This result further emphasizes the practical superiority of our method.

\begin{table}[h]
    \centering
    \caption{
        Results of the user study. We report the average score, the number of users who rate each method as the best on average across all samples (\#Pref. users), and the number of samples for which each method is rated the best on average across all users (\#Win samples).
    }
    \begin{tabular}{lccc}
        \toprule
        & Score & \#Pref. users & \#Win samples\\
        \midrule
        LayerD & \bf{3.74} & \bf{15 (71.4\%)} & \bf{27 (54.0\%)} \\
        YOLO base & 3.52 & 2 (9.5\%) & 15 (30.0\%) \\
        VLM base & 3.31 & 4 (19.0\%) & 8 (16.0\%) \\
        \bottomrule
    \end{tabular}
    \label{tab:user-study}
\end{table}

\section{Influence of Matting and Inpainting Model Choices}
We vary the matting backbones (Swin-L/T~\cite{liu2021swin}, PVT-M/S~\cite{wang2021pyramid}) and replace the inpainting model with FLUX.1 Fill [dev]~\cite{blackforest2024fluxfill} and evaluate their influence.
The larger matting models improve performance while using FLUX.1 Fill [dev] shows significant degradation.
Generative inpainting often introduces unwanted objects, which interfere with subsequent decomposition steps.
This highlights the need for graphic design-specific inpainting as well as refinement.

\begin{figure}[H]
  \centering
  \begin{subfigure}[b]{\linewidth}
    \centering
    \includegraphics[width=\linewidth]{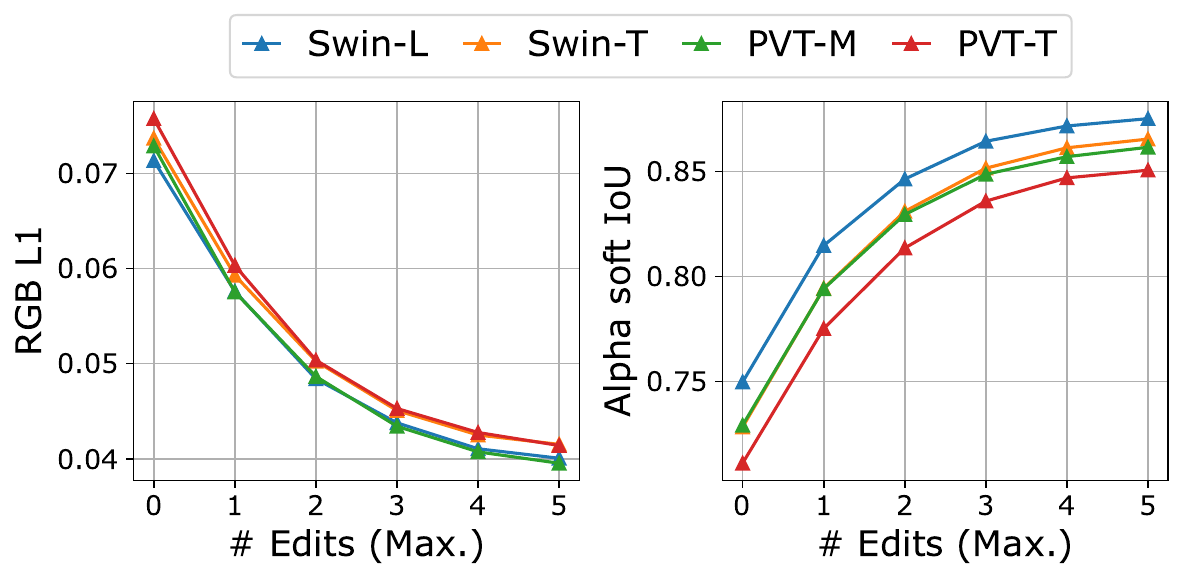}
    \caption{Results with different matting backbones, SwinTransformer~\cite{liu2021swin} and PVT~\cite{wang2021pyramid} variants. The inpainting model is fixed to LaMa~\cite{lama}.}
    \label{fig:matting_ablation}
  \end{subfigure}
  \hfill
  \begin{subfigure}[b]{\linewidth}
    \centering
    \includegraphics[width=\linewidth]{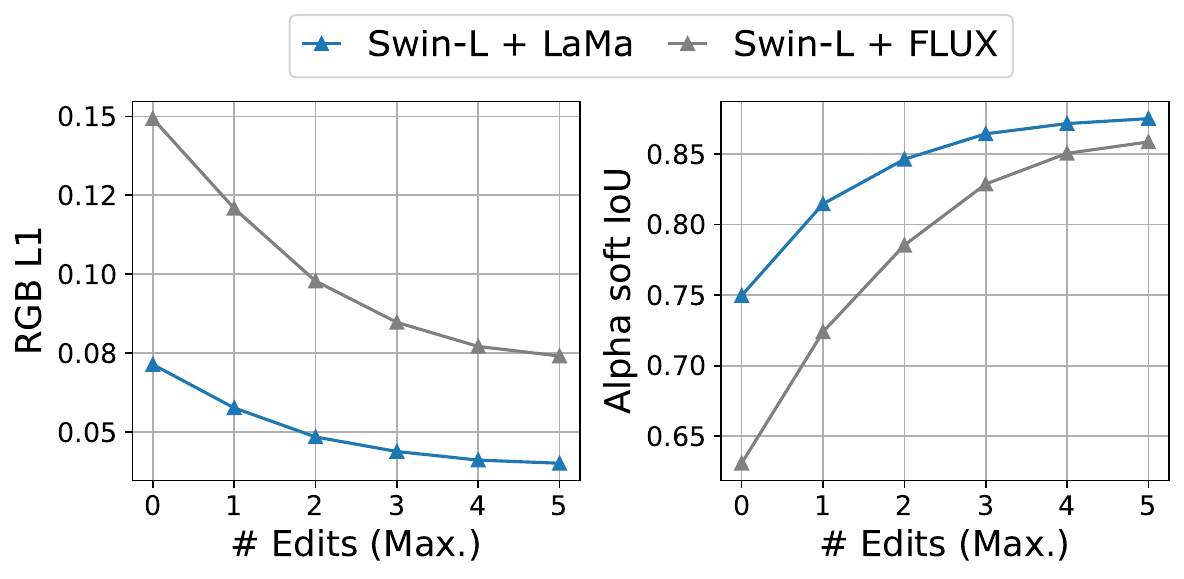}
    \caption{Results with different inpainting models, LaMa~\cite{lama} and FLUX~\cite{flux}.}
    \label{fig:inpaint_ablation}
  \end{subfigure}

  \caption{Evaluation results of \ours{} with different matting (a) and inpainting model (b) choices.}
  \label{fig:ablation}
\end{figure}

\section{Detail of Decomposition Metrics}
\subsection{Dynamic Time Warping}
\label{sup:dtw}
We implement the Dynamic Time Warping (DTW) as shown in \cref{alg:dtw}. 
Given decomposition results $\hat{Y}=(\hat{\bm{l}}_k)_{k=0}^{K}$ and ground truth $Y=(\bm{l}_q)_{q=0}^{Q}$, the output pairs must include $(0,0)$ and $(K,Q)$ as the start point and end point with a step size of 1, and every layer must be included in at least one pair. An average distance is then computed over all pairs as the final output.
\begin{algorithm}[]
\caption{Dynamic Time Warping (DTW)}
\label{alg:dtw}
\definecolor{codeblue}{rgb}{0.25,0.5,0.5}
\definecolor{codekw}{rgb}{0.85, 0.18, 0.50}
\lstset{
  breaklines=true,
  columns=fullflexible,
  basicstyle=\fontsize{7.2pt}{7.2pt}\ttfamily\selectfont,
  commentstyle=\fontsize{7.2pt}{7.2pt}\color{codeblue},
  keywordstyle=\fontsize{7.2pt}{7.2pt}\color{codekw},
}
\begin{lstlisting}[language=python]
# Inputs:
#  - ls: decomposition results of length K (from bottom to top)
#  - gts: ground truth of length Q (from bottom to top)
#  - dist: distance func bounded in [0, 1]
#
# Outputs:
#  - pairs: a list of (l_idx, gt_idx)
#  - D: distance

# Step 1: Compute Cost Matrix
C = np.zeros((len(ls), len(gts))) 
for i in range(len(ls)):
    for j in range(len(gts)):
        C[i,j] = dist(ls[i], ls[j])

# Step 2: Compute Accumulated Cost Matrix
D = np.zeros((len(ls), len(gts))) 
for i in range(1, len(ls)):
    D[i, 0] = D[i-1,0] + C[i,0]
for j in range(1, len(gts)):
    D[0, j] = D[0,j-1] + C[0,j]
for i in range(1, len(ls)):
    for j in range(1, len(gts)):
        D[i,j] = C[i, j] + min(D[i-1,j], D[i,j-1], D[i-1,j-1])

# Step 3: Backtrace to Find Optimal Alignment
i, j = len(ls)-1, len(gts)-1
pairs = [(i,j)]
while True:
    if i==0 and j==0:
        break
    elif i==0:
        pairs.append((i,j-1))
        j-=1
    elif j==0:
        pairs.append((i-1,j))
        i -= 1
    elif D[i-1,j-1]<=D[i-1,j] and D[i-1,j-1]<=D[i,j-1]:
        pairs.append((i-1,j-1))
        i -= 1
        j -= 1
    elif D[i-1,j]<=D[i-1,j-1] and D[i-1,j]<=D[i,j-1]:
        pairs.append((i-1,j))
        i -= 1
    else:
        pairs.append((i,j-1))
        j -= 1

D = sum([Dist(ls[i], gts[j]) for i,j in pairs])/len(pairs)

return pairs, D
\end{lstlisting}
\end{algorithm}

\subsection{Edits algorithm}
\label{sup:edits}
\begin{figure*}[p]
    \centering
    \includegraphics[keepaspectratio, width=0.9\linewidth]{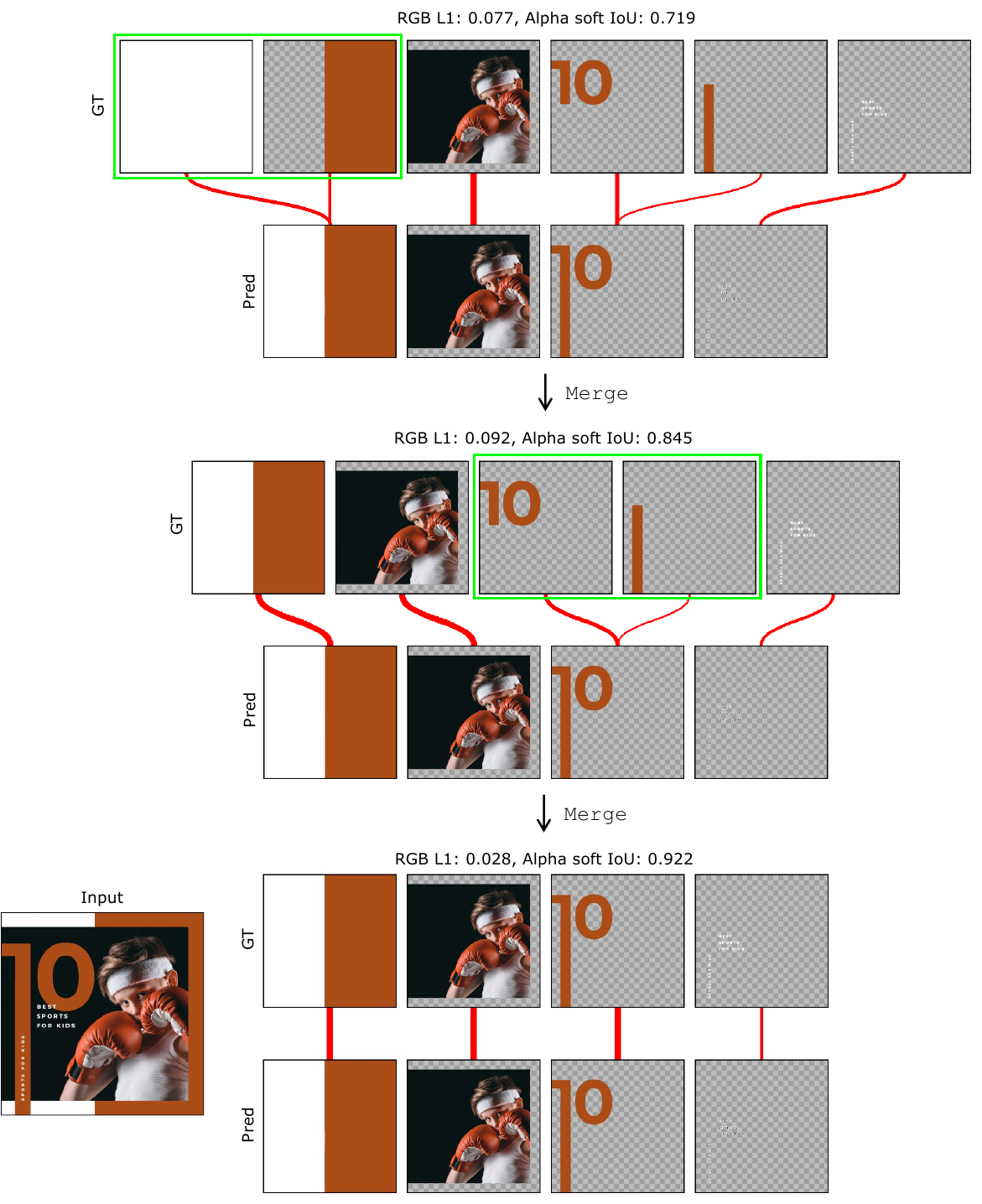}
    \caption{
        Visual example of the DTW-based layer alignment and editing process. Red lines connect matched layers between \ours{}'s prediction and the ground truth; their thickness represents the matching score (the inverse of the distance), \ie, the thicker the line, the higher the score. Green boxes indicate the layers that are merged during the editing process. All layers are sorted from back to front, with the backmost layer on the left and the frontmost on the right.
        Although the decomposition result appears useful for editing the input image, its quality is underestimated due to a mismatch in granularity with the ground truth. Layer merging resolves this mismatch, enabling a more faithful evaluation of the decomposition quality.
    }
    \label{fig:sup-edit-process-1}
\end{figure*}

\begin{figure*}[p]
    \centering
    \includegraphics[keepaspectratio, width=0.93\linewidth]{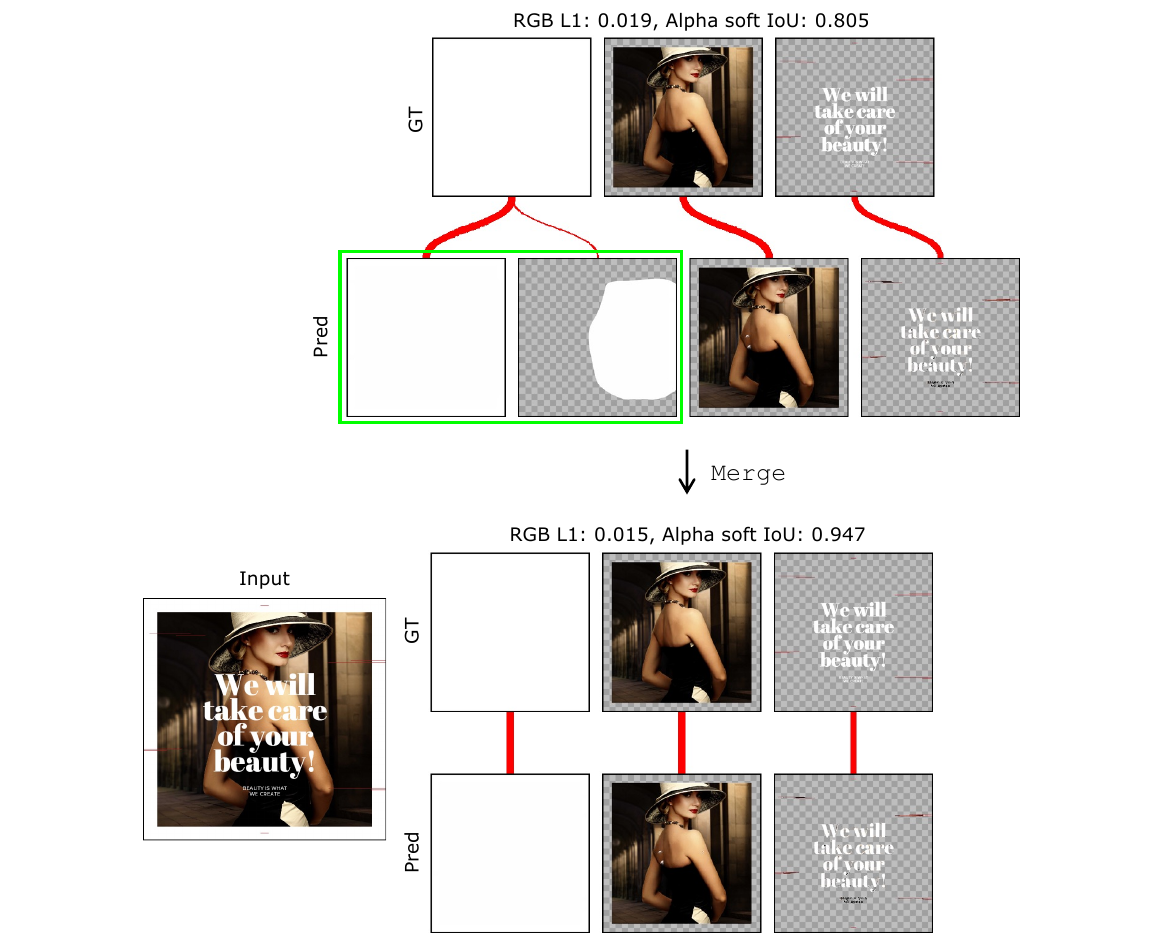}
    \caption{
        Visual example of the DTW-based layer alignment and editing process. Red lines connect matched layers between \ours{}'s prediction and the ground truth; their thickness represents the matching score (the inverse of the distance), \ie, the thicker the line, the higher the score. Green boxes indicate the layers that are merged during the editing process. All layers are sorted from back to front, with the backmost layer on the left and the frontmost on the right.
        \ours{} overdecomposes the white background, but in practical scenarios, it is easy to merge these into a single layer.
        Our evaluation treats such cases as requiring a single edit operation, reflecting the actual editing workload for users.
    }
    \label{fig:sup-edit-process-2}
\end{figure*}

We employ an iterative refinement process with DTW to quantify the number of edits required to align the decomposition results with the given ground truth. At each iteration, we apply the edit (\texttt{Merge}) that yields the highest gain until either the maximum number of edits is reached or the number of layers is reduced to two, as shown in \cref{alg:merge_edit,alg:find_merge_gains}. 
To efficiently approximate the optimal edit, we adopt a greedy search strategy: at iteration $i$, we focus on changes in distances between consecutive layers---specifically, layers $i$, $i+1$, and $i+2$ (if present)---rather than evaluating all layers globally. The optimal edit is then selected from among all candidates at each iteration, ensuring a balance between computational efficiency and alignment accuracy. 
Although \cref{alg:merge_edit,alg:find_merge_gains} describe only the merging of predicted layers for simplicity, we apply the same merging procedure to both the predicted and ground truth layers to address both under- and over-decomposition.
See \cref{fig:sup-edit-process-1,fig:sup-edit-process-2} for visualization of the alignment and merging process.

\begin{algorithm}[h]
\caption{MergeEdit}
\label{alg:merge_edit}
\definecolor{codeblue}{rgb}{0.25,0.5,0.5}
\definecolor{codekw}{rgb}{0.85, 0.18, 0.50}
\lstset{
  breaklines=true,
  columns=fullflexible,
  basicstyle=\fontsize{7.2pt}{7.2pt}\ttfamily\selectfont,
  commentstyle=\fontsize{7.2pt}{7.2pt}\color{codeblue},
  keywordstyle=\fontsize{7.2pt}{7.2pt}\color{codekw},
}
\begin{lstlisting}[language=python]
# Inputs:
#  - ls: decomposition results of length K (bottom to top)
#  - gts: ground truth of length Q (bottom to top)
#  - emax: maximum number of edits 
#  - dist: distance function bounded in [0, 1]
#
# Outputs:
#  - pairs: a list of (l_idx, gt_idx)
#  - D: distance
#  - e: number of edits

e = 0
while e < emax and len(ls) > 2: 
    pairs, _ = dtw(ls, gts)
    merged_ids, gains = find_gains(ls, gts, 
                                pairs, dist)
    if len(gains) > 0:
        best_id = merged_ids[argmin(gains)]
        merged = merge(ls[best_id], ls[best_id+1])
        ls[best_id] = merged
        ls.pop(best_id+1)
    else:
        break
    e += 1
return dtw(ls, gts), e


def merge(x, y):  # Merge func by OpenCV 
    return Image.alpha_composite(x, y)
\end{lstlisting}
\end{algorithm}

\clearpage

\begin{algorithm}[h]
\caption{FindGains}
\label{alg:find_merge_gains}
\definecolor{codeblue}{rgb}{0.25,0.5,0.5}
\definecolor{codekw}{rgb}{0.85, 0.18, 0.50}
\lstset{
  breaklines=true,
  columns=fullflexible,
  basicstyle=\fontsize{7.2pt}{7.2pt}\ttfamily\selectfont,
  commentstyle=\fontsize{7.2pt}{7.2pt}\color{codeblue},
  keywordstyle=\fontsize{7.2pt}{7.2pt}\color{codekw},
}
\begin{lstlisting}[language=python]
# Inputs:
#  - ls: decomposition results of length K (bottom to top)
#  - gts: ground truth of length Q (bottom to top)
#  - pairs: list of (l_idx, gt_idx) obtained from DTW
#  - dist: distance function bounded in [0, 1]
#
# Outputs:
#  - merged_ids: list of indices where merging occurs
#  - gains: list of corresponding distance reductions

merged_ids, gains = [], []
for i in range(len(ls)-1):
    # Step 1: Compute merged layer candidates
    subls = [merge(ls[i], ls[i+1])] + ([ls[i+2]] if i+2 < len(ls) else [])
    
    # Step 2: Gather corresponding ground truth layers
    subgts = [
        [gts[p[1]] for p in pairs if p[0] == i],
        [gts[p[1]] for p in pairs if p[0] == i+1]
    ]

    # Step 3: Compute current distance sum
    curD = sum([dist(ls[i], subgt) for subgt in subgts[0]]) + \
            sum([dist(ls[i+1], subgt) for subgt in subgts[1]])

    # Step 4: Compute distance sum after merging
    Ds = []
    for j in range(len(subls)):
        for k in range(len(subgts)):
            Ds.append(sum([dist(subls[j], subgt) for subgt in subgts[k]]))      
    Ds = [d + Ds[0] for d in Ds[1:]]
    minD = min(Ds)        
    
    # Step 5: Check if merging reduces distance
    if minD < curD:
        merged_ids.append(i)
        gains.append(minD - curDs)
return merged_ids, gains


def merge(x, y):  # Merge func by OpenCV 
    return Image.alpha_composite(x, y)
\end{lstlisting}
\end{algorithm}

\newpage

\section{Loss functions}
\label{sup:loss}

We use binary cross-entropy loss $\mathcal{L}_{\text{BCE}}$, IoU loss $\mathcal{L}_{\text{IoU}}$, and SSIM loss $\mathcal{L}_{\text{SSIM}}$ in our training as BiRefNet~\citep{birefnet}. Definitions of each loss function are as follows.
\begin{align}
    \mathcal{L}_{\text{BCE}}(\hat{\bm{l}}^{\text{A}}, \bm{l}^{\text{A}}) &= \frac{1}{|\Omega|} \sum_{i,j\in\Omega} -\bm{l}^{\text{A}}_{i,j} \log \hat{\bm{l}}^{\text{A}}_{i,j} \notag \\
    &\quad - (1 - \bm{l}^{\text{A}}_{i,j}) \log (1 - \hat{\bm{l}}^{\text{A}}_{i,j}),
\end{align}
\begin{equation}
    \mathcal{L}_{\text{IoU}}(\hat{\bm{l}}^{\text{A}}, \bm{l}^{\text{A}}) = 1 -  \frac{\sum\limits_{i,j\in\Omega}\bm{l}^{\text{A}}_{i,j} \hat{\bm{l}}^{\text{A}}_{i,j}}{\sum\limits_{m,n\in\Omega}\bm{l}^{\text{A}}_{m,n} + \hat{\bm{l}}^{\text{A}}_{m,n} - \bm{l}^{\text{A}}_{m,n} \hat{\bm{l}}^{\text{A}}_{m,n}},
\end{equation}
{\scriptsize
\begin{equation}
    \mathcal{L}_{\text{SSIM}}(\hat{\bm{l}}^{\text{A}}, \bm{l}^{\text{A}}) = 1- \frac{1}{|\mathcal{P}|} \sum_{p\in \mathcal{P}} \frac{(2\mu_{\bm{l}^{\text{A}}_p} \mu_{\hat{\bm{l}}^{\text{A}}_p} + C_1)(2\sigma_{\bm{l}^{\text{A}}_p\hat{\bm{l}}^{\text{A}}_p} + C_2)}{(\mu_{\bm{l}^{\text{A}}_p}^2 + \mu_{\hat{\bm{l}}^{\text{A}}_p}^2 + C_1)(\sigma_{\bm{l}^{\text{A}}_p}^2 + \sigma_{\hat{\bm{l}}^{\text{A}}_p}^2 + C_2)},
\end{equation}
}
where $\Omega$ denotes the set of spatial indices, and $\mathcal{P}$ represents the set of overlapping patches. The local mean $\mu_{\hat{\bm{l}}^{\text{A}}_p}$ and variance $\sigma^2_{\hat{\bm{l}}^{\text{A}}_p}$, as well as the local mean $\mu_{\bm{l}^{\text{A}}_p}$ and variance $\sigma^2_{\bm{l}^{\text{A}}_p}$ of ground truth, are computed within corresponding patches indexed by $p \in \mathcal{P}$. The covariance $\sigma_{\bm{l}^{\text{A}}_p\hat{\bm{l}}^{\text{A}}_p}$ quantifies structural similarity between the prediction and ground truth patches. $C_1$ and $C_2$ are constants and the setting details follow \cite{birefnet}, except that both the predicted and ground-truth alpha maps $\hat{\bm{l}}^{\text{A}}$ and $ \bm{l}^{\text{A}}$,  are not binarized due to shading and smooth transitions commonly used in graphic design.

\end{document}